\documentclass[preprint,aps,12pt,showpacs,nofootinbib,tightenlines]{revtex4}
\usepackage{mathrsfs}
\usepackage{amsmath}
\usepackage{amssymb}
\usepackage{epsfig}
\usepackage{graphicx}
\newcommand{\psl}{ P \hspace{-2.4truemm}/ }
\newcommand{\nsl}{ n \hspace{-2.2truemm}/ }
\newcommand{\vsl}{ v \hspace{-2.2truemm}/ }
\newcommand{\esl}{ \epsilon \hspace{-1.5truemm}/ }

\def\be{\begin{eqnarray}}
\def\en{\end{eqnarray}}
\def\non{\nonumber\\}

\def\prd{{Phys. Rev. D}~}
\def\prl{{ Phys. Rev. Lett.}~}
\def\plb{{ Phys. Lett. B}~}
\def\npb{{ Nucl. Phys. B}~}

\newcommand{\acp}{{\cal A}_{CP}}
\begin{document}
\title{Perturbative  QCD for $B_s \to a_1(1260)(b_1(1235))P(V)$ Decays}
\author{Zhi-Qing Zhang
\footnote{Electronic address: zhangzhiqing@haut.edu.cn} } 
\affiliation{\it \small  Department of Physics, Henan University of
Technology, Zhengzhou, Henan 450052, P.R.China } 
\date{\today}
\begin{abstract}
Within the framework of perturbative QCD approach, we study the charmless two-body decays $B_s \to a_1(1260)(b_1(1235))P(V)$
($P, V$ represent the light pseudo-scalar and vector mesons, respectively.). Using the decays constants and the light-cone distribution amplitudes for these mesons derived from the QCD sum rule
method, we find the following results: (a) The decays $\bar B^0_s\to  a^{-}_1K^{+}(K^{*+})$ have the
contributions from the factorization emission diagrams with a large
Wilson coefficient $C_2+C_1/3$ (order of 1), so they have the largest
branching ratios and arrive at $10^{-5}$ order. While for the decays
$\bar B^0_s\to  a^{0}_1 K^{0}( K^{*0})$, the Wilson
coefficient is $C_1+C_2/3$ in tree level and color suppressed, so
their branching ratios are small and fall in the order of
$10^{-7}\sim10^{-8}$. For the decays $\bar B^0_s\to b_1K(K^*)$, all of their branching ratios are of order few times $10^{-6}$. (b) For the pure annihilation type decays $\bar B^0_s\to a_1(b_1)\rho$ except the decays $\bar B^0_s\to a_1\pi$ having large branching
ratios of order few times $10^{-6}$, the most other decays have the branching ratios of $10^{-7}$ order.
The branching ratios of the decays $\bar B^0_s\to a^0_1(b^0_1)\omega$ are the smallest and fall in the order of $10^{-8}\sim10^{-9}$.
(c)The branching ratios and the direct CP-asymmetries of decays $\bar B^0_s\to a^0_1(b_1^0)\eta^{(\prime)}$ are very sensitive to take
different Gegenbauer moments for $\eta^{(\prime)}$. (d) Except for the decays $\bar B^0_s\to  a^{0}_1 K^{*0}, a^{0}_1\omega, b^{0}_1\omega$, the
longitudinal polarization fractions of other $\bar B^0_s\to  a_1(b_1)V$ decays are very large and more than $90\%$.
(e) Compared with decays $\bar B^0_s\to a_1(b_1)P$, most of $\bar B^0_s\to a_1(b_1)V$ decays have smaller direct CP asymmetries.
\end{abstract}

\vspace{1cm}
\maketitle

\section{Introduction}\label{intro}
In general, the mesons are classified in $J^{PC}$ multiplets. There
are two types of orbitally excited axial-vector mesons, namely
$1^{++}$ and $1^{+-}$. The former includes $a_1(1260), f_1(1285),
f_1(1420)$ and $K_{1A}$, which compose the $^3P_1$-nonet, and the
latter includes $b_1(1235), h_{1}(1170), h_1(1380)$ and $K_{1B}$,
which compose the $^1P_1$-nonet. There is an important character for
these axial-vector mesons except $a_1(1260)$ and $b_1(1235)$, that
is each different flavor state can mix with one another, which comes
from the other nonet meson or the same nonet one.

$B^0\to a^{\pm}_1(1260)\pi^{\mp}$ are the first decay modes with an axial-vector in the
final state observed by BarBar and Belle \cite{aubert, abe, aubert2}. Measuring their time-dependent CP asymmetries can provide the information of Cabibbo-Kobayshi-Maskawa (CKM) weak phase
$\alpha$. After these measurements, many other charmless decays $B\to AP, AV$ ($P, V$ stand for the light pseudo-scalar and vector
mesons) have also been reported by experiments \cite{aubert3,blanc,burke,abe2,babar,babar2,babar3}.
On the theoretical side, many methods are employed to research these decays, such as the naive factorization approach\cite{laporta,CMV},
the generalized factorization approach \cite{chen}, the QCD factorization approach \cite{cheng1, cheng2}, the PQCD approach \cite{wwang}. Though the
factorization approach holds only approximately and its predictions are at odds with experiments for some decays, many factorization
approaches can explain the data in many cases. So these results predicted by the different factorization approaches are useful to
investigate production mechanism of axial vectors in $B$ meson decays, extract the information of Cabibbo-Kobayshi-Maskawa (CKM) weak phase,
probe the structures of axial vectors, even calculate the relative strong phase between tree and penguin diagrams. To our knowledge there is
still lacking the study of charmless decays $B_s\to AP, AV$ both in experiments and theories. Our aim is to fill in this gap and provide
a ready reference to the forthcoming experiments to compare their data with the predictions in the PQCD approach.
In view of the fact that $a_1(1260)$ and $b_1(1235)$ can not mix with each other because of the opposite
C-parities and they do not also mix with other mesons, we would like to study the decays $\bar B_s\to a_1(1260)P(V), b_1(1235)P(V)$ in detail.

In the following, $a_1(1260)$ and $b_1(1235)$ are denoted as $a_1$ and $b_1$ in some places for
convenience. The layout of this paper is as follows. In Sec.\ref{proper}, decay constants and light-cone
distribution amplitudes of the relevant mesons are introduced. In Sec.\ref{results}, we then analyze these decay channels using the PQCD
approach. The numerical results and the discussions are given in
Sec. \ref{numer}. The conclusions are presented in the final part.


\section{decay constants and distribution amplitudes }\label{proper}

For the wave function of the heavy $B_s$ meson,
we take
\be
\Phi_{B_s}(x,b)=
\frac{1}{\sqrt{2N_c}} (\psl_{B_s} +m_{B_s}) \gamma_5 \phi_{B_s} (x,b).
\label{bmeson}
\en
Here only the contribution of Lorentz structure $\phi_{B_s} (x,b)$ is taken into account, since the contribution
of the second Lorentz structure $\bar \phi_{B_s}$ is numerically small \cite{cdlu} and has been neglected. For the
distribution amplitude $\phi_{B_s}(x,b)$ in Eq.(\ref{bmeson}), we adopt the following model:
\be
\phi_{B_s}(x,b)=N_{B_s}x^2(1-x)^2\exp[-\frac{M^2_{B_s}x^2}{2\omega^2_b}-\frac{1}{2}(\omega_bb)^2],
\en
where $\omega_b$ is a free parameter, we take $\omega_b=0.5\pm0.05$ Gev in numerical calculations, and $N_{B_s}=63.671$
is the normalization factor for $\omega_b=0.4$.

The wave functions for the pseudo-scalar (P) mesons $K, \pi$ are
given as \be \Phi_{P}(P,x,\zeta)\equiv
\frac{1}{\sqrt{2N_C}}\gamma_5 \left [ \psl \phi^{A}(x)+m_0
\phi^{P}(x)+\zeta m_0 (\vsl \nsl - v\cdot
n)\phi^{T}(x)\right ],  \en
where the parameter $\zeta$ is either $+1$ or $-1$ depending on the
assignment of the momentum fraction $x$. The chiral scale parameter $m_0$ is defined as $m_0=\frac{M^2_P}{m_{q_1}+m_{q_2}}$.
The distribution amplitudes are expanded as:
\be
\phi^A_{K, \pi}(x)&=&\frac{3f_{K, \pi}}{\sqrt{6}}x(1-x)\left[1+a_{1(K, \pi)}C^{3/2}_1(t)+a_{2(K, \pi)}C^{3/2}_2(t)\right],\label{kpi}\\
\phi^p_{K}(x)&=&\frac{3f_{K}}{2\sqrt{6}}\left[1+0.43C^{1/2}_{2}(t)\right];\phi^p_{\pi}(x)=\frac{3f_{\pi}}{2\sqrt{6}}\left[1+0.24C^{1/2}_{2}(t)\right],\\
\phi^T_K(x)&=&\frac{-f_K}{2\sqrt{6}}\left[C^{1/2}_1(t)+0.35C^{1/2}_3(t)\right];\phi^T_\pi(x)=\frac{-f_\pi}{2\sqrt{6}}\left[C^{1/2}_1(t)+0.55C^{1/2}_3(t)\right],
\en
with Gegenbauer polynomials defined as:
\be
C^{3/2}_1(t)&=&3t, C^{3/2}_2(t)=1.5(5t^2-1),\\
C^{1/2}_1(t)&=&t, C^{1/2}_{2}(t)=0.5(3t^2-1), C^{1/2}_3(t)=0.5t(5t^2-3). \label{kpi1}
\en
As for the distribution amplitudes of the pseudo-scalar mesons $\eta$ and $\eta^\prime$, we use the quark flavor basis mixing mechanism proposed by Refs.\cite{feldmann} and take the same formulae and parameter values as those in Ref.\cite{zjxiao}.

For the vector mesons, their
distribution amplitudes are defined as
\be
\langle V(P, \epsilon^*_L)|\bar q_{2\beta}(z)q_{1\alpha}(0)|0\rangle&=&\frac{1}{\sqrt{2N_c}}\int^1_0dx \; e^{ixp\cdot z}[m_V\esl^*_L\phi_V(x)+\esl^*_L \psl\phi_V^{t}(x)
+m_V\phi^{s}_V(x)]_{\alpha\beta},\non
\langle V(P, \epsilon^*_T)|\bar q_{2\beta}(z)q_{1\alpha}(0)|0\rangle&=&\frac{1}{\sqrt{2N_c}}\int^1_0dx \; e^{ixp\cdot z}\left[m_V\esl^*_T\phi^v_V(x)+\esl^*_T \psl\phi_V^{T}(x)
\right.\non && \left.+m_Vi\epsilon_{\mu\nu\rho\sigma}\gamma_5\gamma^\mu\epsilon^{*v}_Tn^\rho v^\sigma\phi^{a}_V(x)\right]_{\alpha\beta},
\en
where $n (v)$ is the unit vector having the same (opposite) direction with the moving of the vector meson and  $x$ is the momentum fraction of
$q_2$ quark. The distribution amplitudes of the axial-vectors have the same format as those of the vectors except the factor $i\gamma_5$ from the left hand:
\be
\langle A(P, \epsilon^*_L)|\bar q_{2\beta}(z)q_{1\alpha}(0)|0\rangle&=&\frac{i\gamma_5}{\sqrt{2N_c}}\int^1_0dx \; e^{ixp\cdot z}[m_A\esl^*_L\phi_A(x)+\esl^*_L \psl\phi_A^{t}(x)
+m_A\phi^{s}_A(x)]_{\alpha\beta},\non
\langle A(P, \epsilon^*_T)|\bar q_{2\beta}(z)q_{1\alpha}(0)|0\rangle&=&\frac{i\gamma_5}{\sqrt{2N_c}}\int^1_0dx \; e^{ixp\cdot z}\left[m_A\esl^*_T\phi^v_A(x)+\esl^*_T \psl\phi_A^{T}(x)
\right.\non && \left.+m_Ai\epsilon_{\mu\nu\rho\sigma}\gamma_5\gamma^\mu\epsilon^{*v}_Tn^\rho v^\sigma\phi^{a}_A(x)\right]_{\alpha\beta}.
\en
\begin{table}
\caption{Decay constants and Gegenbauer moments for each meson (in MeV). The values are taken at $\mu=1$ GeV.}
\begin{center}
\begin{tabular}{|c|c|c|c|}
\hline \hline  $f_{K^*}$ & $f^T_{K^*}$ &$f_\phi$&$f^T_\phi$\\
 $209\pm2$ &$165\pm9$&$231\pm4$&$186\pm9$\\
 \hline $f_K$& $f_{\pi}$&$f_{a_1}$&$f^T_{b_1}$ \\
 $160$&$130$&$238\pm10$ &$-180\pm8$\\
\hline  $f_\rho$ & $f^T_\rho$ & $f_\omega$& $f^T_\omega$\\
 $209\pm2$ &$165\pm9$&$195\pm3$&$151\pm9$\\
\hline
$a_{1K}$&$a_{1\pi}$&$a_{2K}$&$a_{2\pi}$\\
$0.17$&$0$&$0.2$&$0.44$\\
\hline
$a^\parallel_1(K^*)$&$a^\perp_1(K^*)$&$a^\parallel_2(K^*)$&$a^\perp_2(K^*)$\\
$0.03\pm0.02$&$0.04\pm0.03$&$0.11\pm0.09$&$0.10\pm0.08$\\
\hline
$a^\parallel_2(\rho,\omega)$&$a^\perp_2(\rho,\omega)$&$a^\parallel_2(\phi)$&$a^\perp_2(\phi)$\\
$0.15\pm0.07$&$0.14\pm0.06$&$0.18\pm0.08$&$0.14\pm0.07$\\
\hline
$a^\parallel_2(a_1(1260))$&$a^\perp_1(a_1(1260))$&$a^\parallel_1(b_1(1235))$&$a^\perp_2(b_1(1235))$\\
$-0.02\pm0.02$&$-1.04\pm0.34$&$-1.95\pm0.35$&$0.03\pm0.19$\\
\hline\hline
\end{tabular}\label{gegen}
\end{center}
\end{table}
As for the upper twist-2 and twist-3 distribution functions of the final state mesons, $\phi_{V(A)}$, $\phi_{V(A)}^t$, $\phi_{V(A)}^s$, $\phi^T_{V(A)}$,
$\phi^v_{V(A)}$ and $\phi^a_{V(A)}$ can be calculated by using the light-cone QCD sum rule. We list the distribution functions of the vector (V) mesons, namely
$\rho(\omega, \phi)$, as follows
\be
\begin{cases}
\phi_V(x)=\frac{f_{V}}{2\sqrt{2N_c}}\phi_\parallel(x), \phi^T_V(x)=\frac{f^T_{V}}{2\sqrt{2N_c}}\phi_\perp(x),\\
\phi^t_V(x)=\frac{f^T_V}{2\sqrt{2N_c}}h^{(t)}_\parallel(x), \phi^s_V(x)=\frac{f^T_V}{2\sqrt{4N_c}}\frac{d}{dx}h^{(s)}_\parallel(x),\\
\phi^v_V(x)=\frac{f_V}{2\sqrt{2N_c}}g^{(v)}_\perp(x), \phi^a_V(x)=\frac{f_V}{8\sqrt{2N_c}}\frac{d}{dx}g^{(a)}_\perp(x). \label{vamp}
\end{cases}
\en
The axial-vector (A) mesons , here $a_1$ and $b_1$, can be obtained by replacing each $\phi_V$ with $\phi_A$, by replacing
$f^T_V(f_V)$ with $f$ in Eq.(\ref{vamp}). Here we use $f$ to present both longitudinally and transversely polarized mesons $a_1(b_1)$ by assuming $f^T_{a_1}=f_{a_1}=f$ for $a_1$
and $f_{b_1}=f^T_{b_1}=f$ for $b_1$. In Eq.(\ref{vamp}), the twist-2 distribution functions are in the first line and can be expanded as
\be
\phi_{\parallel,\perp}&=&6x(1-x)\left[1+a^{\parallel,\perp}_2\frac{3}{2}(5t^2-1)\right], \quad\quad\quad\quad\quad\quad\mbox{ for $V$ mesons};\label{t1} \\
\phi_{\parallel,\perp}&=&6x(1-x)\left[a^{\parallel,\perp}_0+3a^{\parallel,\perp}_1t+a^{\parallel,\perp}_2\frac{3}{2}(5t^2-1)\right], \quad\mbox{ for $A$ mesons},
\en
where the zeroth Gegenbauer moments $a^{\perp}_0(a_1)=a^{\parallel}_0(b_1)=0$ and $a^{\parallel}_0(a_1)=a^{\perp}_0(b_1)=1$.

As for twist-3 LCDAs, we use the asymptotic forms for $V$ mesons:
\be
h^{(t)}_\parallel(x)&=&3t^2, h^{(s)}_\parallel(x)=6x(1-x),\non
g^{(a)}_\perp(x)&=&6x(1-x), g^{(v)}_\perp(x)=\frac{3}{4}(1+t^2).
\en
And we use the following forms for $A$ mesons:
\be
h^{(t)}_\parallel(x)&=&3a^\perp_0t^2+\frac{3}{2}a^\perp_1t(3t^2-1), h^{(s)}_\parallel(x)=6x(1-x)(a^\perp_0+a^\perp_1t),\non
g^{(a)}_\perp(x)&=&6x(1-x)(a^\parallel_0+a^\parallel_1t), g^{(v)}_\perp(x)=\frac{3}{4}a^\parallel_0(1+t^2)+\frac{3}{2}a^\parallel_1t^3. \label{t4}
\en
In Eqs.(\ref{kpi})-(\ref{kpi1}) and Eqs.(\ref{t1})-(\ref{t4}), the function $t=2x-1$. The decays constants and the Gegenbauer moments $a^{\parallel,\perp}_n$ for each meson are quoted the numerical
results \cite{camsler, pball1, pball2, pball3, yang1, yang2} and listed in Table \ref{gegen}.


\section{ the perturbative QCD  calculation} \label{results}
The PQCD approach is an effective theory to handle hadronic $B_s$ decays. Because it takes into account the transverse momentum of the valence
quarks in the hadrons, one will encounter double logarithm divergences when the soft and the collinear momenta overlap. Fortunately, these large
double logarithm can be re-summed into the Sudakov factor \cite{hnli0}. There are also another type of double logarithms which arise from the loop corrections
to the weak decay vertex. These double logarithms can also be re-summed and resulted in the threshold factor \cite{hnli00}. This factor decreases faster than any other
power of the momentum fraction in the threshold region, which removes the endpoint singularity. This factor is often parameterized
into a simple form which is independent on channels, twists and flavors \cite{hnli}. Certainly, when the higher order diagrams only suffer from soft or
collinear infrared divergence, it is ease to cure by using the eikonal approximation \cite{hnli2}. Controlling these
kinds of divergences reasonably makes the PQCD approach more self-consistent.

\begin{figure}[t,b]
\vspace{-3cm} \centerline{\epsfxsize=16 cm \epsffile{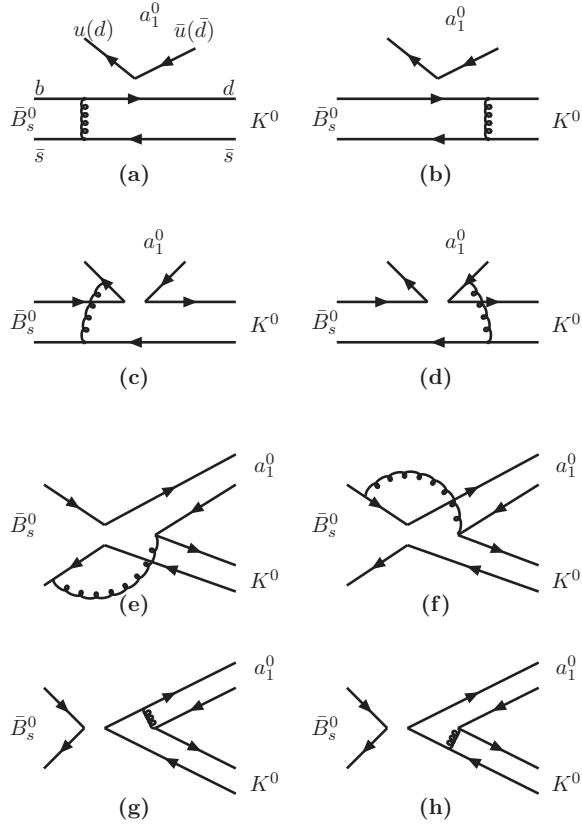}}
\vspace{-9cm} \caption{ Diagrams contributing to the decay $\bar B_s^0\to a^0_1K^0$.}
 \label{fig1}
\end{figure}

Here we take the decay $\bar B^0_s\to a^0_1K^0$ as an example, whose all of the single hard gluon exchange diagrams are shown in Figure 1.
These diagrams contain all of the leading order contributions to the decay $\bar B^0_s\to a^0_1K^0$
in the PQCD approach. Diagrams 1(a) and 1(b) are called factorizable emission diagrams, where $a_1(1250)$ is at emitted position, the corresponding amplitude is presented as
$F_{eK}$. In the PQCD approach, the form factor can be extracted from this amplitude. If $a_1(1250)$ is replaced with $b_1(1235)$, the amplitude contributed by the $(V-A)(V\pm A)$
operators would be zero due to the vanishing decay constant $f_{b_1}$. Diagrams 1(c) and 1(d) are called nonfactorizable emission diagrams, the corresponding amplitude is
represented as $M_{ek}$. For the decay $\bar B^0_s\to a^0_1K^0$, $F_{ek}$ included the color suppressed Wilson coefficient $C_2/3+C_1$ gives the dominated contributions,
while for the decay $\bar B^0_s\to b^0_1K^0$, $M_{ek}$ included the color allowed Wilson coefficient $C_2$ gives the dominated contribution. Diagrams 1(e), 1(f) and 1.(g),
1.(h) are called nonfactorizable and factorizable annihilation diagrams, respectively, and the corresponding amplitudes are written as $M_{ak}$ and $F_{ak}$.
For the decays $\bar B^0_s\to a_1(b_1)\pi$, only these annihilation type amplitudes can contribute to the final results. If the meson $K^0$ is replaced with
a vector meson $K^*(\rho, \omega, \phi)$, the amplitudes will become complicated, for both longitudinal and transverse polarizations can contribute to the decay width.
So we can get three kinds of polarization
amplitudes $M_L$ (longitudinal) and $M_{N,T}$ (transverse) by calculating these diagrams. Because of the aforementioned distribution amplitudes of the axial-vectors
having the same format as those of the vectors except a factor, so the formulas of here considered $\bar B^0_s\to a_1(b_1)V$ decays can be obtained from the ones
of $\bar B^0_s\to VV$ decays by some replacements.
\section{Numerical results and discussions} \label{numer}

We use the following input parameters in the numerical calculations \cite{pdg10,ckmfit}:
\be
f_{B_s}&=&230 MeV, M_{B_s}=5.37 GeV, M_W=80.41 GeV,\\
\tau_{B_s}&=&1.472\times 10^{-12} s, \alpha=91.0^\circ, \gamma=67.2^\circ,\\
|V_{td}|&=&8.58\times10^{-3}, |V_{ts}|=0.03996,|V_{tb}|=0.999,\\
|V_{ud}|&=&0.97425,  |V_{us}|=0.22539, |V_{ub}|=3.54\times10^{-3}.
\en

In the $B_s$-rest frame, the decay rates of $\bar B_s\to a_1(b_1)V$, where $V$ represents $K^*, \rho, \omega, \phi$,
can be written as
\be \Gamma=\frac{G_F^2(1-r^2_{a_1(b_1)})}{32\pi M_B}\sum_{\sigma=L,N,T}{\cal M}^{\sigma\dagger}{\cal M}^{\sigma}, \en
where ${\cal M}^{\sigma}$ is the total decay amplitude
of each considered decay. The subscript $\sigma$ is the helicity states of the two final
mesons with one longitudinal component and two transverse ones. The decay amplitude can be decomposed into three scalar amplitudes $a, b, c$
according to
\be
{\cal M}^{\sigma}&=&\epsilon^*_{2\mu}(\sigma)\epsilon^*_{3\nu}(\sigma)\left[a g^{\mu\nu}+\frac{b}{M_2M_3}P^\mu_BP^\nu_B+
i\frac{c}{M_2M_3}\epsilon^{\mu\nu\alpha\beta} P_{2\alpha}P_{3\beta}\right]\non
&=&{\cal M}_L+{\cal M}_N\epsilon^*_2(\sigma=T)\cdot\epsilon^*_3(\sigma=T)+i\frac{{\cal M}_T}{M^2_B}\epsilon^{\alpha\beta\gamma\rho}
\epsilon^*_{2\alpha}(\sigma)\epsilon^*_{3\beta}(\sigma)P_{2\gamma}P_{3\rho},
\en
where $M_2$ and $M_3$ are the masses of the two final mesons $a_1(b_1)$ and $K^*(\rho,\omega,\phi)$, respectively. The amplitudes
${\cal M}_L, {\cal M}_N, {\cal M}_T$ can be
expressed as
\be
{\cal M}_L&=&a \;\epsilon^*_2(L)\cdot\epsilon^*_3(L)+\frac{b}{M_2M_3}\epsilon^*_{2}(L)\cdot P_3\epsilon^*_{3}(L)\cdot P_2,\non
{\cal M}_N&=&a, \;\;\;\;{\cal M}_T=\frac{M^2_B}{M_2M_3}c.
\en
We can use the amplitudes with different Lorentz structures to define the helicity amplitudes, one longitudinal amplitudes $H_0$
and two transverse amplitudes $H_{\pm}$:
\be
H_0=M^2_B{\cal M}_L,\;\;\; H_{\pm}=M^2_B{\cal M}_N\mp M_2M_3\sqrt{r^2-1}{\cal M}_T,
\label{ampde} \en
where the ratio $r=P_2\cdot P_3/(M_2M_3)$. After the helicity summation, we can get the relation
\be
\sum_{\sigma=L,N,T}{\cal M}^{\sigma\dag}{\cal M}^\sigma=|{\cal M}_L|^2+2\left(|{\cal M}_N|^2+|{\cal M}_T|^2\right)=|H_0|^2+|H_+|^2+|H_-|^2.
\en

The matrix elements ${\cal M}_{j}$ of the operators in the weak Hamilitonian can be calculated by using PQCD approach,
which are written as
as
\be
M_j&=&V_{ub}V^*_{ud(s)}T_{j}-V_{tb}V^*_{td(s)}P_{j}\non
&=&V_{ub}V^*_{ud(s)}T_{j}(1+z_je^{i(\alpha(\gamma)+\delta_j)}), \label{am}
\en
where $j=L, N, T$ and $\alpha$ and $\gamma$ are the Cabibbo-Kobayashi-Maskawa
weak phase angles, defined via $\alpha=arg[-\frac{V_{tb}V^*_{td}}{V_{ub}V^*_{ud}}]$ and $\gamma=arg[-\frac{V_{tb}V^*_{ts}}{V_{ub}V^*_{us}}]$, respecitvely.
$\delta_{j}$ is the relative strong phase between
the tree and the penguin amplitudes, which are denoted as "$T_{j}$" and
"$P_{j}$", respectively. The term $z_{j}$ describes the ratio of penguin to
tree contributions and is defined as \be
z_j=\left|\frac{V_{tb}V^*_{td(s)}}{V_{ub}V^*_{ud(s)}}\right|\left|\frac{P_j}{T_j}\right|.
\en
In the same way, it is easy to write decay amplitude
$\overline {\cal M}_j$ for the corresponding conjugated decay mode:
\be
\overline {\cal M}_j&=&V^*_{ub}V_{ud(s)}T_{j}-V^*_{tb}V_{td(s)}P_{j}\non
&=&V^*_{ub}V_{ud(s)}T_{j}(1+z_je^{i(-\alpha(\gamma)+\delta_j)}).\label{amcon}
\en
So the CP-averaged branching ratio for each considered decay is defined
as \be {\cal B}=(|{\cal M}_j|^2+|\overline{\cal
M}_j|^2)/2&=&|V_{ub}V^*_{ud(s)}|^2\left[T^2_L(1+2z_L\cos\alpha(\gamma)\cos\delta_L+z_L^2)\right.\non &&\left.
+2\sum_{j=N,T}T^2_j(1+2z_j\cos\alpha(\gamma)\cos\delta_j+z_j^2)\right].\label{brann}
\en
Like the decays $\bar B^0_s\to VV$, there are also $3$ types of helicity amplitudes, so corresponding to $3$ types of $z_j$ and $\delta_j$,
respectively. Compared with the decays $\bar B^0_s\to a_1(b_1)V$, the calculation formula for the branching ratios of other considered
decay modes
$\bar B^0_s\to a_1(b_1)P$ are simpler, for only the
longitudinal polarized component of the axial-vector combining with the distribution amplitudes of the pseudo-scalar meson can
contribute to the final branching ratio.

Using the input parameters and the wave functions as specified in this section and Sec.\ref{proper}, it is easy to get the branching ratios for
the considered decays which are listed in Table \ref{bran}, where the first two errors are the $B_s$ wave function shape parameter $\omega_b=0.5\pm0.05$ GeV and
the $B_s$ meson decay constant $f_{B_s}=0.23\pm0.02$ GeV, respectively. The third error is induced by the hard scale-dependent varying from $\Lambda^{(5)}_{QCD}=0.25\pm0.05$
GeV. The last error is from threshold resummation parameter $c$, varying from $0.3$ to $0.4$. The dominant topologies contributing to these
decays are also indicated through the symbols $T$(tree), $P$(penguin), $P_{EW}$(electroweak penguins), $C$(color-suppressed tree) and $ann$
(annihilation).

\begin{table}
\caption{ Branching ratios (in units of $10^{-6}$) for the decays
$\bar B^0_s\to a_1(b_1)K(\pi, \eta, \eta^{\prime})$ and $\bar B^0_s\to
a_1(b_1)K^*(\rho, \omega, \phi)$. In our results, the errors for
these entries correspond to the uncertainties from the $B_s$ meson wave function shape parameter $\omega_B$,  the $B_s$ meson decay constant $f_{B_s}$,
the QCD scale $\Lambda^{(5)}_{QCD}$ and the
threshold resummation parameter $c$, respectively.}
\begin{center}
\begin{tabular}{c|c|c|c|c|c|c|c}
\hline\hline   & Class  & Br$(10^{-6})$&& Class  & Br$(10^{-6})$ \\
\hline
$\bar B^0_s\to  a^{0}_1 K^{0}$ &$C$&$0.081^{+0.016+0.005+0.013+0.029}_{-0.010-0.005-0.011-0.029}$&$\bar B^0_s\to  a^{0}_1 K^{*0}$ &$C$&$0.69^{+0.19+0.03+0.10+0.12}_{-0.13-0.04-0.12-0.12}$\\
$\bar B^0_s\to  a^{-}_1K^{+}$ &$T$&$21.4^{+8.1+0.1+0.9+7.0}_{-5.5-0.0-1.5-7.0}$&$\bar B^0_s\to  a^{-}_1K^{*+}$ &$T$&$29.4^{+10.3+0.1+0.6+9.8}_{-7.2-0.1-1.8-9.8}$\\
$\bar B^0_s\to  a^{-}_1\pi^+$ &$ann$&$2.7^{+0.7+0.2+0.3+0.0}_{-0.5-0.1-0.4-0.0}$&$\bar B^0_s\to  a^{-}_1\rho^+$ &$ann$&$0.38^{+0.3+0.1+0.5+0.8}_{-0.3-0.1-0.7-0.8}$\\
$\bar B^0_s\to  a^{+}_1\pi^-$ &$ann$&$1.8^{+0.5+0.0+0.2+0.0}_{-0.4-0.1-0.3-0.0}$&$\bar B^0_s\to  a^{+}_1\rho^-$ &$ann$&$0.37^{+0.2+0.1+0.3+0.4}_{-0.5-0.1-0.7-0.4}$\\
$\bar B^0_s\to  a^{0}_1\pi^0$ &$ann$&$2.2^{+0.7+0.1+0.4+0.0}_{-0.4-0.0-0.2-0.0}$&$\bar B^0_s\to  a^{0}_1\rho^0$ &$ann$&$0.38^{+0.3+0.1+0.5+0.6}_{-0.2-0.1-0.6-0.6}$\\
$\bar B^0_s\to  a^{0}_1\eta$ &$P_{EW}$&$0.12^{+0.04+0.00+0.00+0.03}_{-0.04-0.00-0.02-0.03}$&$\bar B^0_s\to  a^{0}_1\omega$ &$ann$&$0.0049^{+0.0003+0.0003+0.0005+0.0004}_{-0.0003-0.0004-0.0002-0.0004}$\\
$\bar B^0_s\to  a^{0}_1\eta^{\prime}$ &$P_{EW}$&$0.30^{+0.09+0.02+0.00+0.10}_{-0.08-0.01-0.03-0.10}$&$\bar B^0_s\to a^{0}_1\phi$ &$P_{EW}$&$0.33^{+0.13+0.00+0.02+0.12}_{-0.08-0.00-0.03-0.12}$\\
\hline
$\bar B^0_s\to  b^{0}_1 K^{0}$ &$T$&$2.8^{+0.5+0.1+0.4+0.1}_{-0.4-0.0-0.3-0.1}$&$\bar B^0_s\to  b^{0}_1 K^{*0}$ &$T$&$3.5^{+0.6+0.1+0.6+0.1}_{-0.5-0.2-0.6-0.1}$\\
$\bar B^0_s\to  b^{-}_1K^{+}$ &$C$&$1.3^{+0.1+0.0+0.1+0.0}_{-0.2-0.1-0.3-0.0}$&$\bar B^0_s\to  b^{-}_1K^{*+}$ &$C$&$2.0^{+0.2+0.1+0.2+0.3}_{-0.2-0.1-0.3-0.3}$\\
$\bar B^0_s\to  b^{-}_1\pi^+$ &$ann$&$0.079^{+0.013+0.001+0.006+0.000}_{-0.013-0.000-0.004-0.000}$&$\bar B^0_s\to  b^{-}_1\rho^+$ &$ann$&$0.88^{+0.06+0.01+0.19+0.05}_{-0.08-0.02-0.18-0.05}$\\
$\bar B^0_s\to  b^{+}_1\pi^-$ &$ann$&$0.17^{+0.02+0.00+0.02+0.00}_{-0.02-0.00-0.01-0.00}$&$\bar B^0_s\to  b^{+}_1\rho^-$ &$ann$&$1.1^{+0.1+0.0+0.2+0.0}_{-0.1-0.0-0.3-0.0}$\\
$\bar B^0_s\to  b^{0}_1\pi^0$ &$ann$&$0.085^{+0.025+0.000+0.002+0.000}_{-0.017-0.000-0.013-0.000}$&$\bar B^0_s\to  b^{0}_1\rho^0$ &$ann$&$0.95^{+0.04+0.01+0.25+0.03}_{-0.06-0.01-0.24-0.03}$\\
$\bar B^0_s\to  b^{0}_1\eta$ &$P_{EW}$&$0.13^{+0.05+0.01+0.01+0.01}_{-0.01-0.01-0.00-0.01}$&$\bar B^0_s\to  b^{0}_1\omega$ &$ann$&$0.011^{+0.001+0.000+0.001+0.002}_{-0.001-0.00-0.000-0.002}$\\
$\bar B^0_s\to  b^{0}_1\eta^{\prime}$ &$P_{EW}$&$0.32^{+0.09+0.02+0.00+0.02}_{-0.04-0.00-0.01-0.02}$&$\bar B^0_s\to b^{0}_1\phi$ &$P_{EW}$&$0.21^{+0.04+0.00+0.03+0.00}_{-0.03-0.00-0.04-0.00}$\\
\hline
\end{tabular}\label{bran}
\end{center}
\end{table}
\subsection{$ \bar B^0_s\to  a_1(b_1)K(K^{*})$}
The decays $\bar B^0_s\to  a^{-}_1K^{+}(K^{*+})$ have the
contributions from the factorization emission diagrams with a large
Wilson coefficient $C_2+C_1/3$ (order of 1), so they have the largest
branching ratios and arrive at $10^{-5}$ order. While for the decays
$\bar B^0_s\to  a^{0}_1 K^{0}( K^{*0})$, the Wilson
coefficient is $C_1+C_2/3$ in tree level and color suppressed, so
their branching ratios are small and fall in the order of
$10^{-7}\sim10^{-8}$. Although the decay $\bar B^0_s\to  a^{0}_1
K^{0}$ is tree dominated, the contributions from tree operators
between the factorization and nonfactorization emission diagrams
cancel each other mostly, which induces its tree amplitudes to have a
very small real part. It does not happen in the channel $\bar
B^0_s\to  a^{0}_1 K^{*0}$. At the same time, there exist three
polarization states for the final mesons and the transverse
polarizations are about $30\%$. So the decay mode $a^{0}_1 K^{*0}$
has a larger branching ratio compared with the mode
$a^{0}_1 K^{0}$. For the decay $\bar B^0_s\to  b^0_1K^0$, the
amplitude of the nonfactorization emission diagrams $M^T_{ek}$ (T denotes the contribution from tree operators) including the large Wilson coefficient $C_2$
receives a larger value, which is about 5 times
the decay $a^0_1K^0$. Furthermore, because of the
vanishing decay constant $f_{b_1}$, the amplitude $F_{eK}$ becomes
zero for the decay $b^0_1K^0$, while which has large value but the
opposite sign with amplitude $M_{ek}$ for the decay $a^0_1K^0$. So
one can find that there is much larger contribution from the tree
operator for the decay $b^0_1K^0$ than that for the decay
$a^0_1K^0$. The decay $\bar B^0_s\to  b^0_1K^{*0}$ has large
branching ratio, which is also because of the large contribution
from the nonfactorizable emission diagrams.
\subsection{$\bar B^0_s\to  a_1(b_1)\pi(\rho,\omega)$}
These channels belong to the annihilation type decays, contributed by the $W-$annihilation and $W-$exchange diagrams. The decays $\bar B^0_s\to  a_1(b_1)\pi(\rho)$ are sensitive to the wave functions of the final states. If the final mesons are $\pi$
and $a_1$, the branching ratios can arrive at $10^{-6}$ order, while for the $\pi$ and $b_1$ final states, the branching ratios become $10^{-7}$ order even smaller.
In a word, ${\cal B}(\bar B^0_s\to a_1\pi)>{\cal B}(\bar B^0_s\to b_1\pi)$. The condition is contrary for the decay modes $a_1(b_1)\rho$. The branching ratios of decays $\bar B^0_s\to \rho^+ a_1^-, \rho^0 a_1^0, \rho^-a_1^+$ are very near each
other. There exists the similar case with the decays $\bar B^0_s\to \rho^+\pi^-, \rho^0\pi^0, \rho^-\pi^+$, whose branching ratios are predicted
as $(2.2,2.3,2.4)\times10^{-7}$ \cite{ali}, respectively. We also show the Cabibbo-Kobayashi-Maskawa angle $\gamma$
dependence of the branching ratios of decays $\bar B^0_s\to a_1(b_1)\pi(\rho)$ in Fig.2. It is easy to see that the branching ratio for the decay
with two neutral mesons in the final state lies the between those of other two decays in most range of $0<\gamma<180^0$.
\begin{figure}[t,b]
\begin{center}
\includegraphics[scale=0.7]{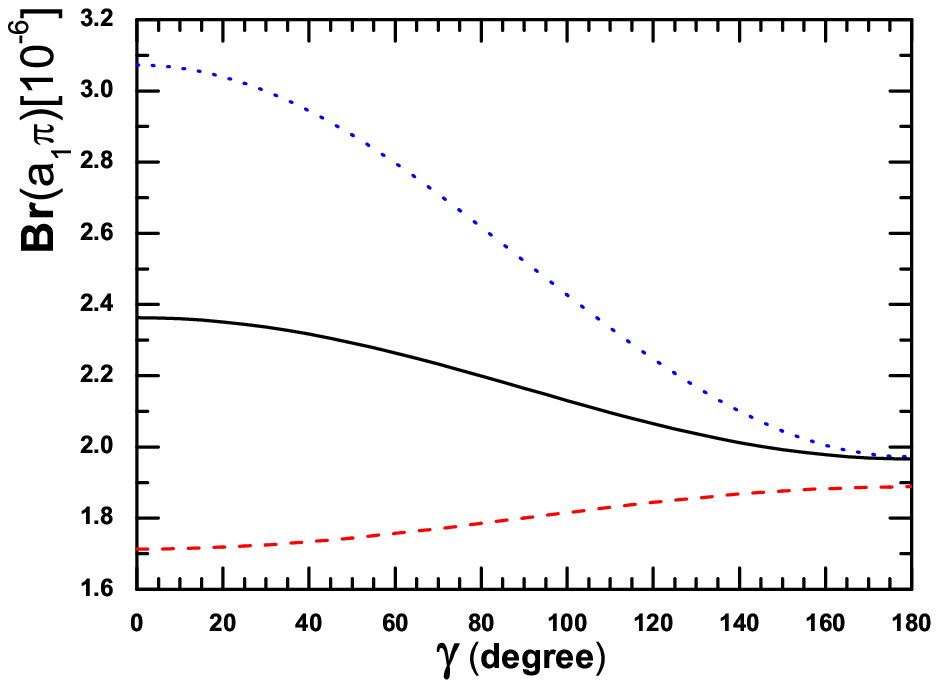}
\includegraphics[scale=0.7]{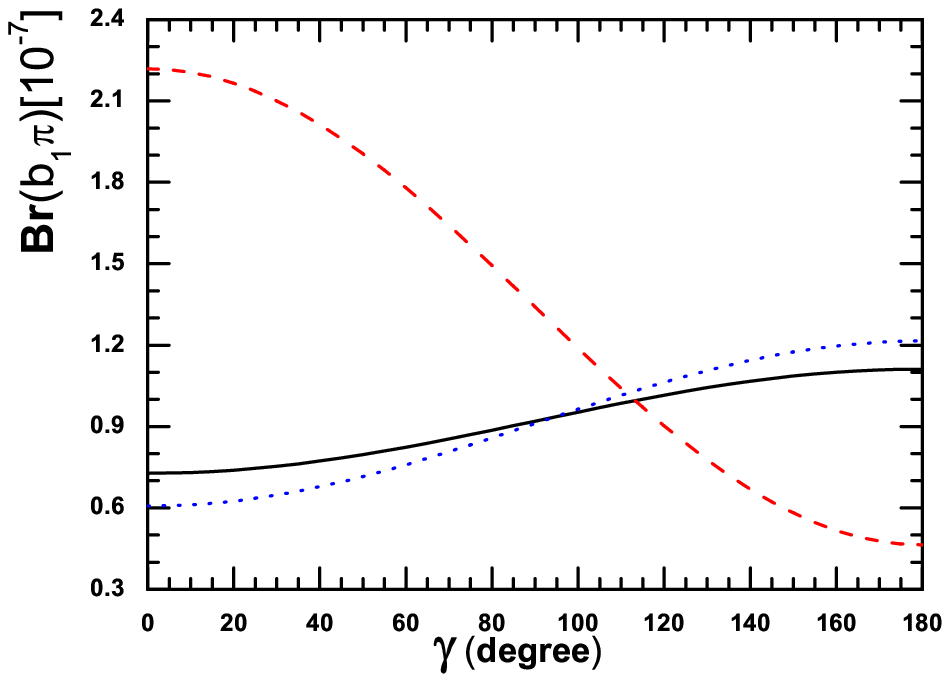}
\includegraphics[scale=0.7]{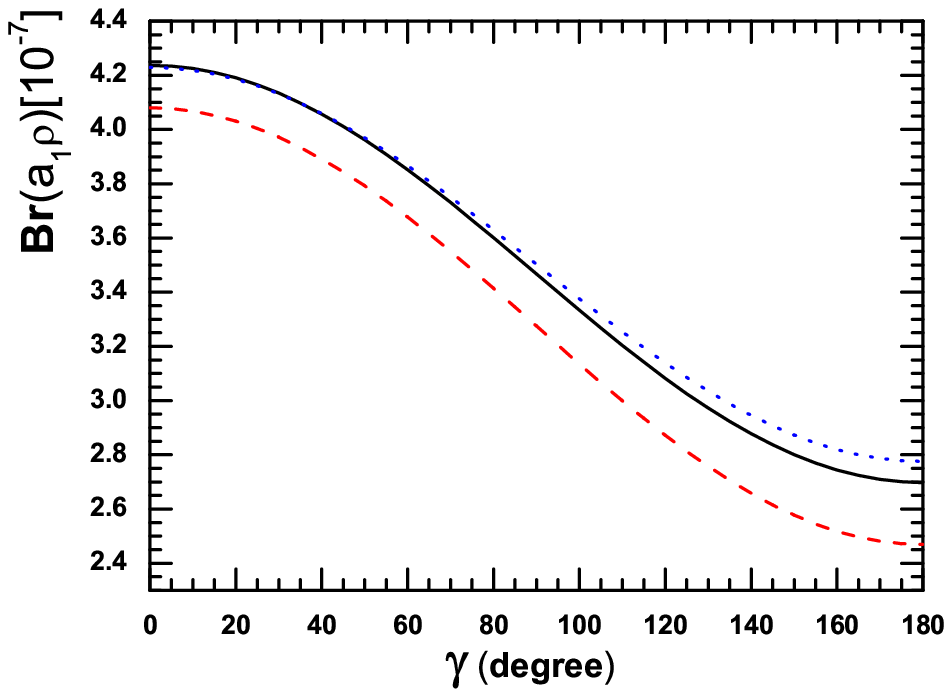}
\includegraphics[scale=0.7]{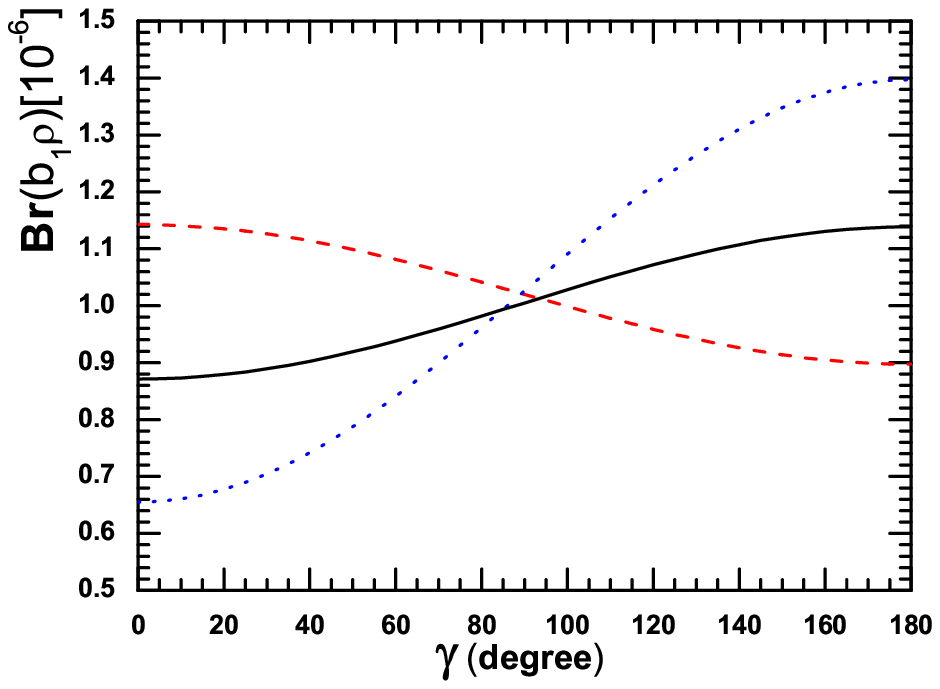}
\vspace{0.3cm} \caption{The dependence of the branching ratios on the
Cabibbo-Kobayashi-Maskawa angle $\gamma$. In these panels, the solid lines are for the decays $\bar B^0_s\to a^{0}_1(b^0_1)\pi^0, a^{0}_1(b^0_1)\rho^0$, dotted lines for
$\bar B^0_s\to a^{-}_1(b^-_1)\pi^+, a^{-}_1(b^-_1)\rho^+$, dashed lines for $\bar B^0_s\to a^{+}_1(b^+_1)\pi^-, a^{+}_1(b^+_1)\rho^-$.}\label{fig2}
\end{center}
\end{figure}

As for the other two annihilation type decays $\bar B^0_s\to  a^{0}_1\omega, b^{0}_1\omega$, whose branching ratios are in the order of $10^{-8}\sim10^{-9}$. It is easy
to see that this kind decay is sensitive to the quark structure of the final mesons. Compared with the decays $\bar B^0_s\to  a^{0}_1(b^0_1)\rho^0$, the difference is
mainly from the signs of $d\bar d$ component in the mesons $\omega$ and $\rho^0$, which induces different interference effects between the amplitudes from the penguin
operators: constructive for the decays $a^{0}_1(b^0_1)\rho^0$, destructive for the decays $a^{0}_1(b^0_1)\omega$. From our calculations, we find that the penguin
amplitude for the decay $a^{0}_1(b^0_1)\rho^0$ is about $20.4(48.2)$ times of that for the decay $a^{0}_1(b^0_1)\omega$.
\subsection{$\bar B^0_s\to  a_1(b_1)\eta^{(\prime)}$}
The main contributions to these four decays are from the electro-weak penguin operators. Although the contributions from the tree operators have a prominent increase
for the decays $\bar B^0_s\to b^0_1\eta^{(\prime)}$ compared with those for the decays $\bar B^0_s\to a^0_1\eta^{(\prime)}$. The former are about $5(7)$ times larger than
the later. For the tree operator contributions are the Cabibbo-Kobayashi-Maskawa suppressed by a factor $50$, so the increased tree operator contributions for the decays
$\bar B^0_s\to b^0_1\eta^{(\prime)}$ bring a slight increase to the branching ratios.

We also checked the sensitivity to the values on the Gegenbauer moments for all the considered decays. If one takes smaller Gegenbauer
moments, such as  $a_1^K=0.05\pm0.02$ \cite{kho}, $0.10\pm0.12$ \cite{braun}, $a_2^{\pi,K}=0.115$ \cite{pball}, the branching ratios have
a few percent change for most of decays $\bar B^0_s\to  a_1(b_1)\pi(K)$, more than 10 percent change for only very few channels. So we considered that the
uncertainties caused by the Gegenbauer moments are small and can be neglected. But it is not the case for the decays $\bar B^0_s\to  a_1(b_1)\eta^{(\prime)}$.
If one takes the newer Gegenbauer
moments as given in Ref. \cite{pball}:
\be
a_2^{\pi}=0.115, a_4^{\pi}=-0.015, \label{newgb}
\en
The branching ratios will have a prominent change,
\be
{\cal B}(\bar B^0_s\to  a^0_1\eta)&=&(0.97^{+0.33+0.01+0.01+0.23}_{-0.34-0.02-0.17-0.23})\times10^{-7},\\
{\cal B}(\bar B^0_s\to  a^0_1\eta^\prime)&=&(2.1^{+0.7+0.0+0.0+0.8}_{-0.5-0.0-0.2-0.8})\times10^{-7},\\
{\cal B}(\bar B^0_s\to  b^0_1\eta)&=&(0.21^{+0.00+0.03+0.05+0.02}_{-0.05-0.05-0.05-0.11})\times10^{-7},\\
{\cal B}(\bar B^0_s\to  b^0_1\eta^\prime)&=&(0.75^{+0.00+0.00+0.16+0.06}_{-0.17-0.16-0.16-0.35})\times10^{-7},
\en
where the errors come from the $B_s$ meson wave function shape parameter $\omega_B=0.5\pm0.05$ GeV,  the $B_s$ meson decay constant $f_{B_s}=0.23\pm0.02$ GeV,
the QCD scale $\Lambda^{(5)}_{QCD}=0.25\pm0.05$ GeV and
threshold resummation parameter $c$ varying from $0.3$ to $0.4$, respectively.
Especially for the decays $\bar B^0_s\to  a^0_1\eta^{(\prime)}$, their branching ratios are sensitive to Gegenbauer moments and increase to
$7\sim8$ times by using
the newer Gegenbauer moments.
Certainly, the increases of the branching ratios for decays $\bar B^0_s\to  b^0_1\eta^{(\prime)}$ are not so large. It is need to clarify which Gegenbauer moments are more reasonable.
\subsection{$\bar B^0_s\to  a_1(b_1)\phi$}
These two decays are dominated by the electro-weak(EW) penguin operators. Though their branching ratios are small, these two decays are interesting
to invest the effect from the electro-weak penguins, where there might exits new physics \cite{buras}. The presence of a new physics contribution from EW can enhance the branching
ratios of the decays $\bar B^0_s\to \pi(\rho)\phi$, which are used to improve the $B\to \pi K$ "puzzle" \cite{hofer}. If here considered
two decays have such effect, it is deserve more research attention.
\section{Polarization fractions of the decays $\bar B^0_s\to  a_1(b_1)V$}
For the decays $\bar B^0_s\to  a_1(b_1)V$, another equivalent set of helicity amplitudes are often used, that is
\be
A_0&=&-M^2_B{\cal M}_L,\non
A_{\parallel}&=&\sqrt{2}M^2_B{\cal M}_N,\non
A_{\perp}&=&M_2M_3\sqrt{2(r^2-1)}{\cal M}_T.
\en
Using this set of helicity amplitudes, we can define three polarization fractions $f_{0, \parallel, \perp}$:
\be
f_{0,\parallel,\perp}=\frac{|A_{0,\parallel,\perp}|^2}{|A_0|^2+|A_\parallel|^2+|A_\perp|^2}.
\en

 \begin{table}
\caption{ Longitudinal polarization fraction ($f_L$) and two transverse polarization fractions ($f_\parallel$, $f_\perp$) for the decays
$\bar B^0_s\to  a_1(b_1)V$. In our results, the uncertainties of $f_L, f_\parallel, f_\perp$ come from the $B_s$ meson wave function shape parameter
$\omega_b$,  the $B_s$ meson decay constant $f_{B_s}$, the QCD scale $\Lambda^{(5)}_{QCD}$ and
threshold resummation parameter $c$, respectively.}
\begin{center}
\begin{tabular}{c|c|c|c|c|c|c|c}
\hline\hline   &   $f_L(\%)$&$f_\parallel(\%)$&$f_\perp(\%)$\\
\hline
$\bar B^0_s\to  a^{0}_1 K^{*0}$ &$68.9^{+6.1+2.7+1.5+3.9}_{-6.4-2.8-2.4-3.9}$&$15.1^{+3.1+1.3+0.9+2.0}_{-3.0-1.2-0.9-2.0}$&$16.0^{+3.4+1.5+1.6+1.9}_{-3.1-1.4-0.8-1.9}$\\
$\bar B^0_s\to  a^{-}_1K^{*+}$ &$90.6^{+0.2+0.3+0.2+0.1}_{-0.3-0.2-0.3-0.1}$&$4.9^{+0.1+0.0+0.2+0.1}_{-0.1-0.1-0.0-0.1}$&$4.5^{+0.1+0.1+0.2+0.1}_{-0.2-0.1-0.1-0.1}$\\
$\bar B^0_s\to  a^{-}_1\rho^+$ &$97.7^{+0.1+0.5+0.6+0.9}_{-0.3-0.4-1.2-0.9}$&$2.2^{+0.2+0.3+1.1+0.8}_{-0.1-0.2-0.6-0.8}$&$0.1^{+0.0+0.0+0.1+0.1}_{-0.0-0.0-0.0-0.1}$\\
$\bar B^0_s\to  a^{+}_1\rho^-$ &$97.8^{+0.2+0.4+0.6+1.0}_{-0.2-0.3-1.1-1.0}$&$2.1^{+0.2+1.1+0.3+1.0}_{-0.2-0.5-0.2-1.0}$&$0.1^{+0.0+0.0+0.0+0.1}_{-0.0-0.0-0.0-0.1}$\\
$\bar B^0_s\to  a^{0}_1\rho^0$ &$97.8^{+0.2+0.3+0.6+1.0}_{-0.1-0.3-1.0-1.0}$&$2.1^{+0.1+0.3+1.0+0.9}_{-0.1-0.3-0.6-0.9}$&$0.1^{+0.0+0.0+0.1+0.1}_{-0.1-0.0-0.1-0.1}$\\
$\bar B^0_s\to  a^{0}_1\omega$ &$83.4^{+1.0+2.4+3.5+6.1}_{-0.9-2.2-2.5-6.1}$&$9.8^{+0.5+1.4+1.0+2.8}_{-0.4-1.3-2.2-2.8}$&$6.8^{+0.3+0.8+1.5+2.3}_{-0.4-0.9-1.4-2.3}$\\
$\bar B^0_s\to  a^{0}_1\phi$ &$94.8^{+0.0+0.1+0.0+0.2}_{-0.0-0.1-0.0-0.2}$&$2.8^{+0.0+0.0+0.0+0.1}_{-0.0-0.0-0.0-0.1}$&$2.4^{+0.0+0.0+0.0+0.1}_{-0.0-0.0-0.1-0.1}$\\
\hline
$\bar B^0_s\to  b^{0}_1 K^{*0}$ &$98.2^{+0.2+0.2+0.2+0.3}_{-0.4-0.2-0.4-0.3}$&$0.9^{+0.1+0.1+0.1+0.1}_{-0.1-0.1-0.1-0.1}$&$0.9^{+0.1+0.2+0.3+0.1}_{-0.1-0.1-0.1-0.1}$\\
$\bar B^0_s\to  b^{-}_1K^{*+}$ &$94.1^{+0.7+0.6+0.8+1.7}_{-0.7-0.6-1.2-1.7}$&$2.8^{+0.3+0.3+0.6+0.8}_{-0.3-0.2-0.4-0.8}$&$3.1^{+0.3+0.3+0.7+0.9}_{-0.4-0.4-0.4-0.9}$\\
$\bar B^0_s\to  b^{+}_1\rho^-$ &$96.9^{+0.3+0.5+0.9+2.7}_{-0.3-0.6-2.3-2.7}$&$2.3^{+0.2+0.4+1.8+2.0}_{-0.2-0.4-1.0-2.0}$&$0.8^{+0.1+0.1+0.5+0.7}_{-0.1-0.1-0.6-0.7}$\\
$\bar B^0_s\to  b^{-}_1\rho^+$ &$91.6^{+0.4+1.5+3.1+4.5}_{-0.6-1.4-5.7-4.5}$&$8.1^{+0.5+1.4+5.5+5.5}_{-0.4-1.2-3.0-5.5}$&$0.3^{+0.1+0.0+0.2+0.1}_{-0.0-0.1-0.1-0.1}$\\
$\bar B^0_s\to  b^{0}_1\rho^0$ &$95.0^{+0.2+0.8+1.9+3.8}_{-0.4-0.9-4.1-3.8}$&$4.7^{+0.3+1.1+3.8+3.6}_{-0.2-1.2-1.8-3.6}$&$0.3^{+0.0+0.1+0.3+0.3}_{-0.0-0.1-0.1-0.3}$\\
$\bar B^0_s\to  b^{0}_1\omega$ &$63.4^{+3.3+3.7+12.7+12.2}_{-2.6-3.8-12.5-12.2}$&$21.7^{+1.7+2.4+7.6+8.2}_{-2.0-2.2-7.5-8.2}$&$14.8^{+1.0+1.5+5.0+4.2}_{-1.2-1.4-5.1-4.2}$\\
$\bar B^0_s\to  b^{0}_1\phi$ &$99.5^{+0.0+0.0+0.0+0.0}_{-0.0-0.0-0.0-0.0}$&$0.25^{+0.01+0.00+0.01+0.00}_{-0.03-0.00-0.03-0.00}$&$0.25^{+0.01+0.00+0.03+0.00}_{-0.01-0.00-0.01-0.00}$\\
\hline\hline
\end{tabular}\label{polar}
\end{center}
\end{table}

The formalism of the wave function has great influence to the polarization fractions for some decays. In Ref.\cite{hnli3}, the author suggested that taking the asymptotic models for
the $K^*$ meson distribution amplitudes instead of its traditional formalism leads to a smaller $B\to K^*$ form factor ($A_0\sim0.3$). The smaller form factor
responds to the smaller longitudinal polarization fraction. Another result is that the strengthened penguin annihilation and nonfactorizable contribuitons
further bring it down. In the decays $\bar B^0_s\to  a_1(b_1) K^{*}$, we also take the asymptotic models for the $K^*$ meson wave functions and only find the decay
mode $a_1^0 K^{*0}$ with smaller longitudinal polarization fraction about $70\%$. If we neglect penguin annihilation contribution in the decay
$\bar B^0_s\to  a^0_1 K^{*0}$, and find that the branching ratio changes from $6.9\times10^{-7}$ to $5.5\times10^{-7}$, while the longitudinal polarization
receives a larger increase and arrives at $93.1\%$.
If we neglect nonfactorizable contribution, both the branching ratio and the polarization fractions will become much smaller.
Compared with $\bar B^0_s\to  a^0_1 K^{*0}$ and $\bar B^0_s\to  b^0_1 K^{*0}$ decays, we argue that
the polarization fractions are also connected with the symmetric properties of $a_1$ and $b_1$ distribution amplitudes, which might have a sensitive effect in the
penguin annihilation contribution. If one neglects penguin annihilation contribution in the decay $\bar B^0_s\to  b^0_1\omega$, the longitude fraction can amount to
$95.4\%$ and the branching ratio decreases by $30\%$. In a word, the contributions from the penguin annihilation diagrams are very sensitive to the final
polarization fractions for some decays.

In Table \ref{polar}, we list the longitudinal polarization fraction ($f_L$) and the transverse polarization fractions ($f_\parallel$, $f_\perp$) for the decays
$\bar B^0_s\to  a_1(b_1)V$, where the errors come from the $B_s$ meson wave function shape parameter $\omega_{b}=0.5\pm0.05$ GeV,  the $B_s$ meson decay constant $f_{B_s}=0.23\pm0.02$ GeV,
the QCD scale $\Lambda^{(5)}_{QCD}=0.25\pm0.05$ GeV and the
threshold resummation parameter $c$ varying from $0.3$ to $0.4$, respectively. Except the decays $\bar B^0_s\to  a^{0}_1 K^{*0}, a^{0}_1\omega, b^{0}_1\omega$, the
longitudinal polarization fractions of other $\bar B^0_s\to  a_1(b_1)V$ decays are very large and more than $90\%$.
\section{Direct CP asymmetry}
Now we turn to the evaluations of the CP-violating asymmetries in PQCD approach. In view that most of $\bar B^0_s\to  a_1(b_1)V$ decays have
small transverse polarization fractions and only about few percent. So we can neglect them in our calculations and the expression for
the direct CP-violating asymmetries of the decays $\bar B^0_s\to  a_1(b_1)V$ (except $\bar B^0_s\to  a^{0}_1 K^{*0}, a^{0}_1\omega, b^{0}_1\omega$)
become simple, which can be got by using Eq.(\ref{am}) and Eq.(\ref{amcon}):
\be
\acp^{dir}&=&\frac{ |\overline{\cal M}|^2-|{\cal M}|^2 }{
 |{\cal M}|^2+|\overline{\cal M}|^2}=\frac{2z_L\sin\alpha(\gamma)\sin\delta_L}
{(1+2z_L\cos\alpha(\gamma)\cos\delta_L+z_L^2) }\;.
\en
The direct CP-violating asymmetries for the decays $\bar B^0_s\to  a_1(b_1)P$ have similar expression.
Using the input parameters and the wave functions as specified in this section and Sec.\ref{proper}, one can calculate the PQCD predictions
(in units of $10^{-2}$) for the direct CP-violating asymmetries of the
considered decays, which are listed in Table \ref{dircp}, where the errors induced by the uncertainties of $\omega_{b}=0.5\pm0.05$ GeV,
$f_{B_s}=0.23\pm0.02$ GeV, $\Lambda^{(5)}_{QCD}=0.25\pm0.05$ GeV and the
threshold resummation parameter $c$ varying from $0.3$ to $0.4$, respectively.
We find the following points:
\begin{table}
\caption{ Direct CP-violating asymmetries (in units of $\%$) for the decays
$\bar B^0_s\to a_1(b_1)K(\pi, \eta, \eta^{\prime})$ and $\bar B^0_s\to a_1(b_1)K^*(\rho, \omega, \phi)$ (except $\bar B^0_s\to  a^{0}_1 K^{*0}, a^{0}_1\omega, b^{0}_1\omega$). In our results, the errors for
these entries correspond to the uncertainties from $\omega_{b}, f_{B_s}$, the QCD scale $\Lambda^{(5)}_{QCD}$ and the
threshold resummation parameter $c$, respectively.}
\begin{center}
\begin{tabular}{c|c|c|c|c|c|c|c}
\hline\hline   & Class  & Br$(10^{-6})$&& Class  & Br$(10^{-6})$ \\
\hline
$\bar B^0_s\to  a^{0}_1 K^{0}$ &$C$&$-66.1^{+10.4+3.3+4.2+37.3}_{-6.5-3.3-5.1-37.3}$&$\bar B^0_s\to  b^{0}_1 K^{0}$ &$T$&$41.4^{+5.3+3.0+2.1+0.3}_{-5.0-3.1-0.8-0.3}$\\
$\bar B^0_s\to  a^{-}_1K^{+}$ &$T$&$-9.7^{+1.4+0.8+0.4+0.7}_{-1.6-0.9-0.2-0.7}$&$\bar B^0_s\to  b^{-}_1K^{+}$ &$C$&$-74.7^{+8.1+3.3+0.4+2.5}_{-7.3-2.6-3.6-2.5}$\\
$\bar B^0_s\to  a^{-}_1\pi^+$ &$ann$&$20.5^{+1.3+0.3+0.4+0.2}_{-1.3-0.0-0.6-0.2}$&$\bar B^0_s\to  b^{-}_1\pi^+$ &$ann$&$12.7^{+2.3+0.0+1.4+0.0}_{-3.3-0.1-2.6-0.0}$\\
$\bar B^0_s\to  a^{+}_1\pi^-$ &$ann$&$3.2^{+0.3+0.0+0.6+0.2}_{-0.3-0.1-0.8-0.2}$&$\bar B^0_s\to  b^{+}_1\pi^-$ &$ann$&$24.5^{+0.7+0.0+2.5+0.1}_{-3.4-0.0-5.3-0.1}$\\
$\bar B^0_s\to  a^{0}_1\pi^0$ &$ann$&$14.0^{+1.1+0.1+0.2+0.0}_{-1.2-0.0-0.0-0.0}$&$\bar B^0_s\to  b^{0}_1\pi^0$ &$ann$&$-23.3^{+2.8+0.1+5.0+0.2}_{-1.3-0.1-2.8-0.2}$\\
$\bar B^0_s\to  a^{0}_1\eta$ &$P_{EW}$&$-31.3^{+0.0+0.3+0.2+4.1}_{-2.8-0.2-5.2-4.1}$&$\bar B^0_s\to  b^{0}_1\eta$ &$P_{EW}$&$25.0^{+0.0+0.0+0.5+3.4}_{-4.0-2.8-4.8-3.4}$\\
$\bar B^0_s\to  a^{0}_1\eta^{\prime}$ &$P_{EW}$&$-10.2^{+1.4+1.2+2.3+2.1}_{-0.0-1.3-0.4-2.1}$&$\bar B^0_s\to b^{0}_1\eta^{\prime}$ &$P_{EW}$&$22.7^{+0.0+0.0+0.9+2.4}_{-3.6-2.5-7.2-2.4}$\\
\hline
--&--&--&$\bar B^0_s\to  b^{0}_1 K^{*0}$ &$T$&$2.7^{+4.2+0.3+5.8+3.2}_{-3.7-0.2-5.2-3.2}$\\
$\bar B^0_s\to  a^{-}_1K^{*+}$ &T&$-11.1^{+1.5+1.0+0.7+1.5}_{-1.7-0.9-0.5-1.5}$&$\bar B^0_s\to  b^{-}_1K^{*+}$ &$C$&$0.80^{+7.4+0.3+7.5+3.9}_{-7.4-0.2-6.6-3.9}$\\
$\bar B^0_s\to  a^{-}_1\rho^+$ &$ann$&$4.3^{+0.6+0.7+1.6+1.5}_{-0.4-0.5-3.3-1.5}$&$\bar B^0_s\to  b^{-}_1\rho^+$ &$ann$&$31.6^{+0.2+0.0+3.6+0.2}_{-0.2-0.1-2.8-0.2}$\\
$\bar B^0_s\to  a^{+}_1\rho^-$ &$ann$&$6.0^{+2.1+1.1+2.4+3.1}_{-0.8-1.0-3.5-3.1}$&$\bar B^0_s\to  b^{+}_1\rho^-$ &$ann$&$-9.3^{+0.2+0.2+0.5+0.0}_{-0.6-0.3-0.4-0.0}$\\
$\bar B^0_s\to  a^{0}_1\rho^0$ &$ann$&$4.6^{+1.3+0.7+1.6+2.1}_{-1.5-0.8-2.9-2.1}$&$\bar B^0_s\to  b^{0}_1\rho^0$ &$ann$&$8.3^{+0.2+0.1+0.8+0.0}_{-0.0-0.2-0.2-0.0}$\\
$\bar B^0_s\to  a^{0}_1\phi$ &$P_{EW}$&$-6.2^{+1.4+0.0+1.4+0.9}_{-1.4-0.0-1.9-0.9}$&$\bar B^0_s\to  b^{0}_1\phi$ &$P_{EW}$&$-0.81^{+0.32+0.00+0.11+0.00}_{-0.15-0.00-0.12-0.00}$\\
\hline\hline
\end{tabular}\label{dircp}
\end{center}
\end{table}
\begin{figure}[t,b]
\begin{center}
\includegraphics[scale=0.7]{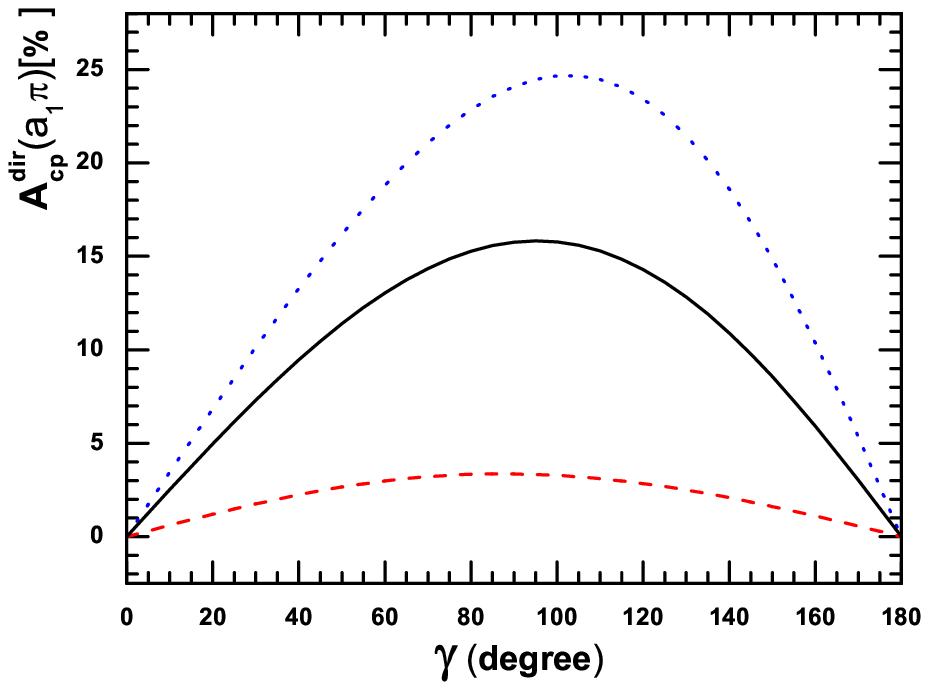}
\includegraphics[scale=0.7]{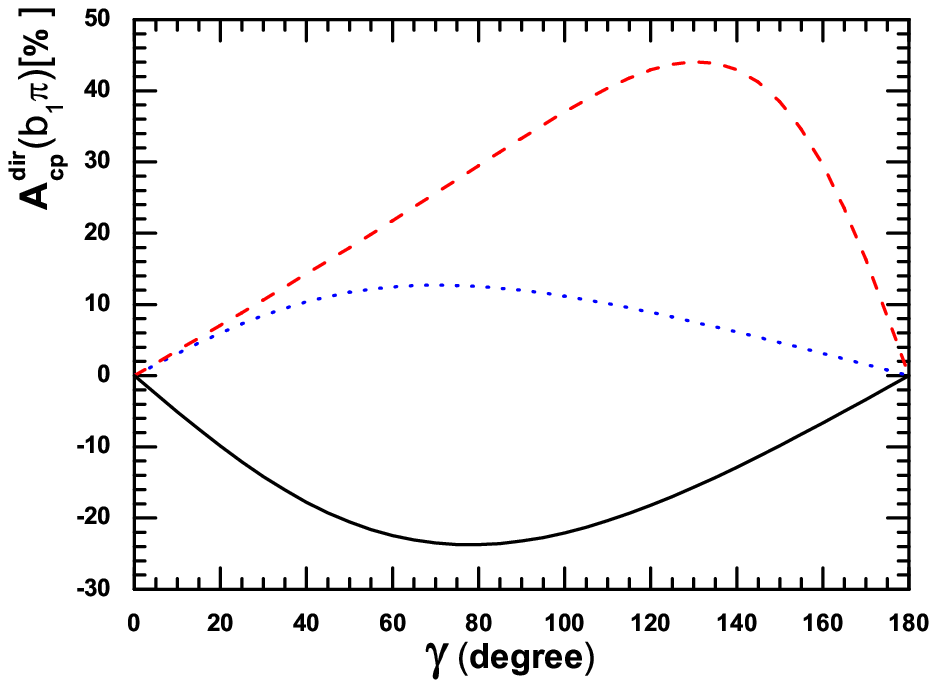}
\includegraphics[scale=0.7]{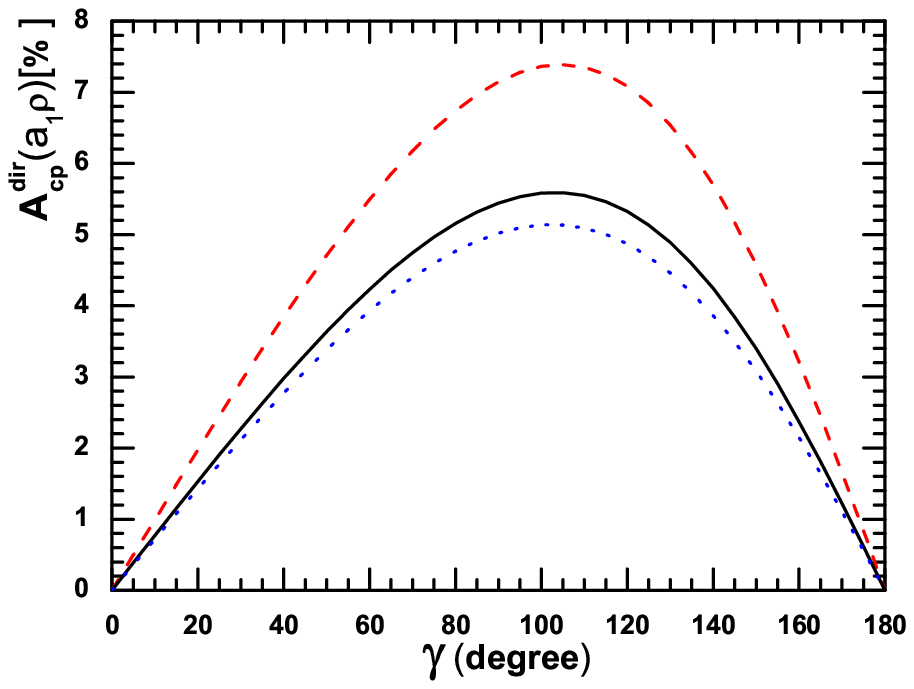}
\includegraphics[scale=0.7]{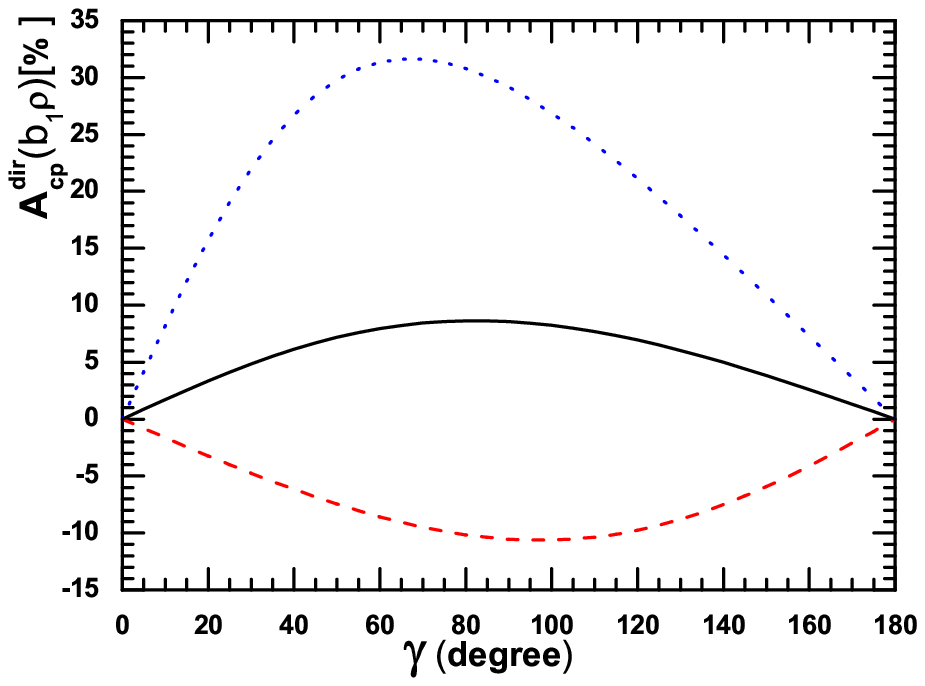}
\vspace{0.3cm} \caption{The dependence of the direct CP-violating asymmetries on the
Cabibbo-Kobayashi-Maskawa angle $\gamma$. In these panels, the solid lines are for the decays $\bar B^0_s\to a^{0}_1(b^0_1)\pi^0, a^{0}_1(b^0_1)\rho^0$, dotted lines for
$\bar B^0_s\to a^{-}_1(b^-_1)\pi^+, a^{-}_1(b^-_1)\rho^+$, dashed lines for $\bar B^0_s\to a^{+}_1(b^+_1)\pi^-, a^{+}_1(b^+_1)\rho^-$.}\label{fig4}
\end{center}
\end{figure}
\begin{itemize}
\item
Like the decay $\bar B^0_s\to  \pi^{0}K^0$, whose direct CP-asymmetry is more than $40\%$ predicted by several methods \cite{ali,beneke2,will}, the decays $\bar B^0_s\to  a^0_1(b^0_1)K^0$
also have large direct CP-asymmetries. Unlike the channel $\bar B^0_s\to  b^{-}_1K^{+}$, the decay $\bar B^0_s\to  a^{-}_1K^{+}$ has a smaller direct CP-asymmetry. It
is because that though there are near penguin amplitudes in theses two decays, the tree amplitude of the latter is about 3 times as large as that of the former,
and the sine values of their strong phases are close to each other. The direct CP-asymmetries in the decays $\bar B^0_s\to a_1(b_1)K^*$ are small.
\item
The direct CP-asymmetries of the decays $\bar B^0_s\to b_1^0\eta^{(\prime)}$ are sensitive to taking different Gegenbauer
moments for $\eta^{(\prime)}$. If we take the newer Gegenbauer moments given in Eq.(\ref{newgb}), their direct CP-asymmetries will change not only in magnitudes but also in signs.
\item
The decays $\bar B^0_s\to a_1(b_1)\rho$ except the channel $\bar B^0_s\to b_1^-\rho^+$ have smaller direct CP-violating asymmetries compared with the decays
$\bar B^0_s\to a_1(b_1)\pi$. The direct CP-violating asymmetry for the decay $\bar B^0_s\to b_1^-\rho^+$ is very sensitive to the tree operator contribution from
the nonfactorization annihilation diagrams: if we neglect such contribution, its branching ratio can increase $14\%$, while the
direct CP-violating asymmetry becomes only $1.3\%$. In Fig.3, we show the dependence of the direct CP-violating asymmetries for the decays
$\bar B^0_s\to a_1(b_1)\pi(\rho)$ on the Cabibbo-Kobayashi-Maskawa angle $\gamma$.
\item
There only exist factorization and nonfactorizaiton emission diagrams for the decays $\bar B^0_s\to a_1(b_1)\phi$. The direct CP-violating asymmetries in these two decays are small,
because the interactions between tree and penguin contributions are small. From our calculations, we find the ratios of penguin to tree amplitudes for decays
$\bar B^0_s\to a_1\phi$ and $\bar B^0_s\to b_1\phi$ are about 0.06 and 0.004, respectively. The strong phases penguin and tree amplitudes are only $0.15$ and $0.026$
rad, respectively.
\item
Compared with decays $\bar B^0_s\to a_1(b_1)P$, most of $\bar B^0_s\to a_1(b_1)V$ decays  have smaller direct CP-violating asymmetries.
\end{itemize}

\section{Conclusion}\label{summary}

In this paper, by using the decay constants and the light-cone distribution amplitudes
derived from QCD sum-rule method, we research  the decays $\bar B^0_s\to a_1(b_1)P, a_1(b_1)V$
in PQCD approach and find that
\begin{itemize}
\item
The decays $\bar B^0_s\to  a^{-}_1K^{+}(K^{*+})$ have the
contributions from the factorization emission diagrams with a large
Wilson coefficient $C_2+C_1/3$ (order of 1), so they have the largest
branching ratios and arrive at $10^{-5}$ order. While for the decays
$\bar B^0_s\to  a^{0}_1 K^{0}( K^{*0})$, the Wilson
coefficient is $C_1+C_2/3$ in tree level and color suppressed, so
their branching ratios are small and fall in the order of
$10^{-7}\sim10^{-8}$. For the decays $\bar B^0_s\to b_1K(K^*)$, all of their branching ratios are of order few times $10^{-6}$.
\item
For the pure annihilation type decays $\bar B^0_s\to a_1(b_1)\rho$ except the decays $\bar B^0_s\to a_1\pi$ having large branching
ratios of order few times $10^{-6}$, the most of them have the branching ratios of $10^{-7}$ order.
The branching ratios of the decays $\bar B^0_s\to a^0_1(b^0_1)\omega$ are the smallest and fall in the order of $10^{-8}\sim10^{-9}$.
\item
The branching ratios and the direct CP-asymmetries of decays $\bar B^0_s\to a^0_1(b_1^0)\eta^{(\prime)}$ are very sensitive to take different Gegenbauer
moments for $\eta^{(\prime)}$.
\item
Except for the decays $\bar B^0_s\to  a^{0}_1 K^{*0}, a^{0}_1\omega, b^{0}_1\omega$, the
longitudinal polarization fractions of other $\bar B^0_s\to  a_1(b_1)V$ decays are very large and more than $90\%$.
\item
Compared with decays $\bar B^0_s\to a_1(b_1)P$, most of $\bar B^0_s\to a_1(b_1)V$ decays have smaller direct CP-violating asymmetries.
\end{itemize}

\section*{Acknowledgment}
This work is partly supported by the National Natural Science Foundation of China under Grant No. 11147004, and by Foundation of Henan University of
Technology under Grant No. 2009BS038.

\end{document}